\documentclass[12pt]{article}

\usepackage{scicite}
\usepackage{graphicx}	% Include figure files

\usepackage{times}

\newenvironment{sciabstract}{%
\begin{quote} \bf}
{\end{quote}}

\newcommand {\lesssim} {\ {\raise-.5ex\hbox{$\buildrel<\over\sim$}}\ }

\newcounter{lastnote}

\title{Pulsar Discovery by Global Volunteer Computing} 

\def\aei{{1}}
\def\LUH{{2}}
\def\UWM{{3}}
\def\cornell{{4}}
\def\NAIC{{5}}
\def\berkeley{{6}}
\def\swinburne{{7}}
\def\mcgill{{8}}
\def\naicithica{{9}}
\def\columbia{{10}}
\def\bonn{{11}}
\def\FMC{{12}}
\def\NRAO{{13}}
\def\UBC{{14}}
\def\stichting{{15}}
\def\AstInst{{16}}
\def\brownsville{{17}}
\def\manchester{{18}}
\def\WVU{{19}}
\def\lafayette{{20}}
\def\aeigolm{{21}}

\author{
B.~Knispel$^{\aei,\LUH, \ast}$,
B.~Allen$^{\aei,\UWM,\LUH}$,
J.~M.~Cordes,$^{\cornell}$,
J.~S.~Deneva,$^{\NAIC}$, \\
D.~Anderson$^\berkeley$,
C.~Aulbert$^{\aei,\LUH}$,
N.~D.~R.~Bhat$^\swinburne$,
O.~Bock$^{\aei,\LUH}$,
S.~Bogdanov$^\mcgill$,\\
A.~Brazier$^{\naicithica, \cornell}$,
F.~Camilo$^\columbia$, 
D.~J.~Champion$^{\bonn}$,
S.~Chatterjee$^\cornell$, 
F.~Crawford$^\FMC$,\\
P.~B.~Demorest$^\NRAO$,
H.~Fehrmann$^{\aei,\LUH}$,
P.~C.~C.~Freire$^\bonn$, 
M.~E.~Gonzalez$^\UBC$,
D.~Hammer$^{\UWM}$,\\
J.~W.~T.~Hessels$^{\stichting, \AstInst}$, 
F.~A.~Jenet$^\brownsville$,
L.~Kasian$^\UBC$, 
V.~M.~Kaspi$^\mcgill$,
M.~Kramer$^{\bonn,\manchester}$,\\
P.~Lazarus$^\mcgill$,
J.~van~Leeuwen$^{\stichting, \AstInst}$,
D.~R.~Lorimer$^{\WVU,\NRAO}$, 
A.~G.~Lyne$^\manchester$
B.~Machenschalk$^{\aei,\LUH}$,\\
M.~A.~McLaughlin$^{\WVU,\NRAO}$,
C.~Messenger$^{\aei,\LUH}$,
D.~J.~Nice$^\lafayette$,
M.~A.~Papa$^{\aeigolm,\UWM}$,
H.~J.~Pletsch$^{\aei,\LUH}$,\\
R.~Prix$^{\aei,\LUH}$,
S.~M.~Ransom$^\NRAO$, 
X.~Siemens$^\UWM$,
I.~H.~Stairs$^\UBC$,
B.~W.~Stappers$^\manchester$,\\
K.~Stovall$^\brownsville$,
A.~Venkataraman$^\NAIC$
\\
\normalsize{$^{\aei}$Albert-Einstein-Institut, Max-Planck-Institut f\"ur Gravitationsphysik, D-30167 Hannover, Germany}\\
\normalsize{$^{\LUH}$Leibniz Universit{\"a}t Hannover, D-30167 Hannover, Germany}\\
\normalsize{$^{\UWM}$Physics Dept., U. of Wisconsin - Milwaukee, Milwaukee WI 53211}\\ 
\normalsize{$^{\cornell}$Astronomy Dept., Cornell Univ., Ithaca, NY 14853}\\ 
\normalsize{$^{\NAIC}$Arecibo Observatory, HC3 Box 53995, Arecibo, PR 00612} \\
\normalsize{$^\berkeley$University of California at Berkeley, Berkeley, CA 94720 USA}\\
\normalsize{$^{\swinburne}$Swinburne U., Center for Astrophysics and Supercomputing, Hawthorn, Victoria 3122, Australia}\\
\normalsize{$^\mcgill$Dept.~of Physics, McGill Univ., Montreal, QC H3A2T8, Canada} \\
\normalsize{$^{\naicithica}$NAIC, Cornell Univ., Ithaca, NY 14853}\\ 
\normalsize{$^{\columbia}$Columbia Astrophysics Laboratory, Columbia Univ., New York, NY 10027}\\
\normalsize{$^\bonn$Max-Planck-Institut f\"ur Radioastronomie, Bonn, Germany} \\
\normalsize{$^{\FMC}$Dept. of Physics and Astronomy, Franklin and Marshall College, Lancaster, PA 17604-3003}\\
\normalsize{$^\NRAO$NRAO (National Radio Astronomy Observatory), Charlottesville, VA 22903} \\
\normalsize{$^\UBC$Dept.~of Physics and Astronomy, Univ.~of British Columbia, Vancouver, BC V6T 1Z1, Canada}\\ 
\normalsize{$^\stichting$Netherlands Inst. for Radio Astronomy (ASTRON), Postbus 2, 7990 AA Dwingeloo, The Netherlands}\\
\normalsize{$^{\AstInst}$Astron. Inst. ``Anton Pannekoek'', Univ. of Amsterdam, 1098 SJ Amsterdam, The Netherlands}\\ 
\normalsize{$^\brownsville$Center for Gravitational Wave Astronomy, U. Texas - Brownsville, TX 78520}\\ 
\normalsize{$^\manchester$Jodrell Bank Centre for Astrophys., School of Phys. and Astr., U. of Manchester, Manch., M13 9PL, UK}\\ 
\normalsize{$^{\WVU}$Dept. of Physics, West Virginia University, Morgantown, WV 26506}\\
\normalsize{$^{\lafayette}$Dept.~of Physics, Lafayette College, Easton, PA 18042} \\
\normalsize{$^{\aeigolm}$Albert-Einstein-Institut, Max-Planck-Institut f\"ur Gravitationsphysik, D-14476 Golm, Germany}\\
\normalsize{$^\ast$Correspondence to benjamin.knispel@aei.mpg.de}
}

\begin{document} 


\baselineskip24pt


\maketitle 
%\date{August 12, 2010}

% PUT PSR J2007+2722 PARAMETERS HERE BEN STAPPERS IS RESPONSIBLE FOR
% THESE VALUES AND ANY LISTED ERRORS.  IF YOU DON'T AGREE, PLEASE
% ARGUE WITH HIM.
%\def\freq{$40.820677596(8)$~Hz }
%\def\freq{$40.820677596 \pm 8 \times 10^{-9}$~Hz }
% Below is DAvid's value, following Ingrid's analysis showing the
% difference between tempo1 and tempo2.  THis is for tempo1.
\def\freq{$40.820677620(6)$~Hz }

%2010:
%40.8206775967685402
% 0.0000000041530968
%PEPOCH        55396.220000


\begin{sciabstract}
% best period as frequency + error
  Einstein@Home aggregates the computer power of hundreds of thousands
  of volunteers from 192 countries to ``mine'' large data sets. It has
  now found a 40.8~Hz isolated pulsar in radio survey data from the
  Arecibo Observatory taken in February 2007. Additional timing
  observations indicate that this pulsar is likely a disrupted
  recycled pulsar.  PSR J2007+2722's pulse profile is remarkably wide
  with emission over almost the entire spin period; the pulsar likely
  has closely aligned magnetic and spin axes. The massive computing
  power provided by volunteers should enable many more such
  discoveries.
  % last sentence might go for length / policy reasons
\end{sciabstract}

Einstein@Home\cite{Einstein@Home} (E@H) is a volunteer distributed
computing project\cite{DistrComp}. Members of the public ``sign up'' their
home or office computers (``hosts''), which automatically download
``workunits'' from the servers, carry out analyses when idle, and
return results. These are automatically validated by comparison with
results for the {\it same} workunit, produced by a different volunteer's
host. More than 250,000 individuals have contributed; each week about
100,000 different computers download work. The aggregate computational
power (0.25 PFlop/s) is on par with the largest supercomputers. E@H's
primary goal is to detect gravitational waves from rapidly-spinning
neutron stars in data from LIGO and VIRGO\cite{Einstein@Home}.

Since 2009, about 35\% of E@H compute cycles have also been used to
search for pulsars in radio data from the PALFA project (see SOM) at
the 305 m Arecibo Telescope.
Data disks are sent to Cornell University's Center for Advanced Computing,
where data are archived. For E@H, data are transferred to
Hannover, de-dispersed for 628 different dispersion measures (DM $\in$ [0, 1002.4]
pc~cm$^{-3}$), and resampled at 128 $\mu s$. Hosts receive
workunits containing time series for four DM values for one beam. Each
is 2MB in size, covering 268.435456~s. A host demodulates each
time series (in the time domain) for 6661 different circular
orbital templates with periods greater than 11 minutes (our Galaxy
has even shorter period binaries). The grid of templates is
spaced so that for pulsar spin-frequencies below 400 Hz, less than
20\% of signal-to-noise ratio is lost. Fourier algorithms sum up to 16
harmonics. A total of 1.85\% of the power spectrum is removed to eliminate
well-known sources of radio frequency interference. A significance
($\mathcal S =- \log_{10} p$, with $p$ the false-alarm probability in
Gaussian noise) is calculated at each grid-point. After $\sim 2$ CPU-hours,
the host uploads the 100 most significant candidates to the server.

When all workunits for a given beam are complete, the results are
post-processed on servers at Hannover. Candidates with $\mathcal S >
15$ are identified by eye, then optimized with PRESTO  (see SOM). To date E@H
has searched 27,000 of 68,000 observed beams. It has re-detected 120
pulsars, most in the past four months, because code and algorithm
optimizations sped-up the search by a factor $\sim 7$.

\begin{figure}
\includegraphics[width=\textwidth]{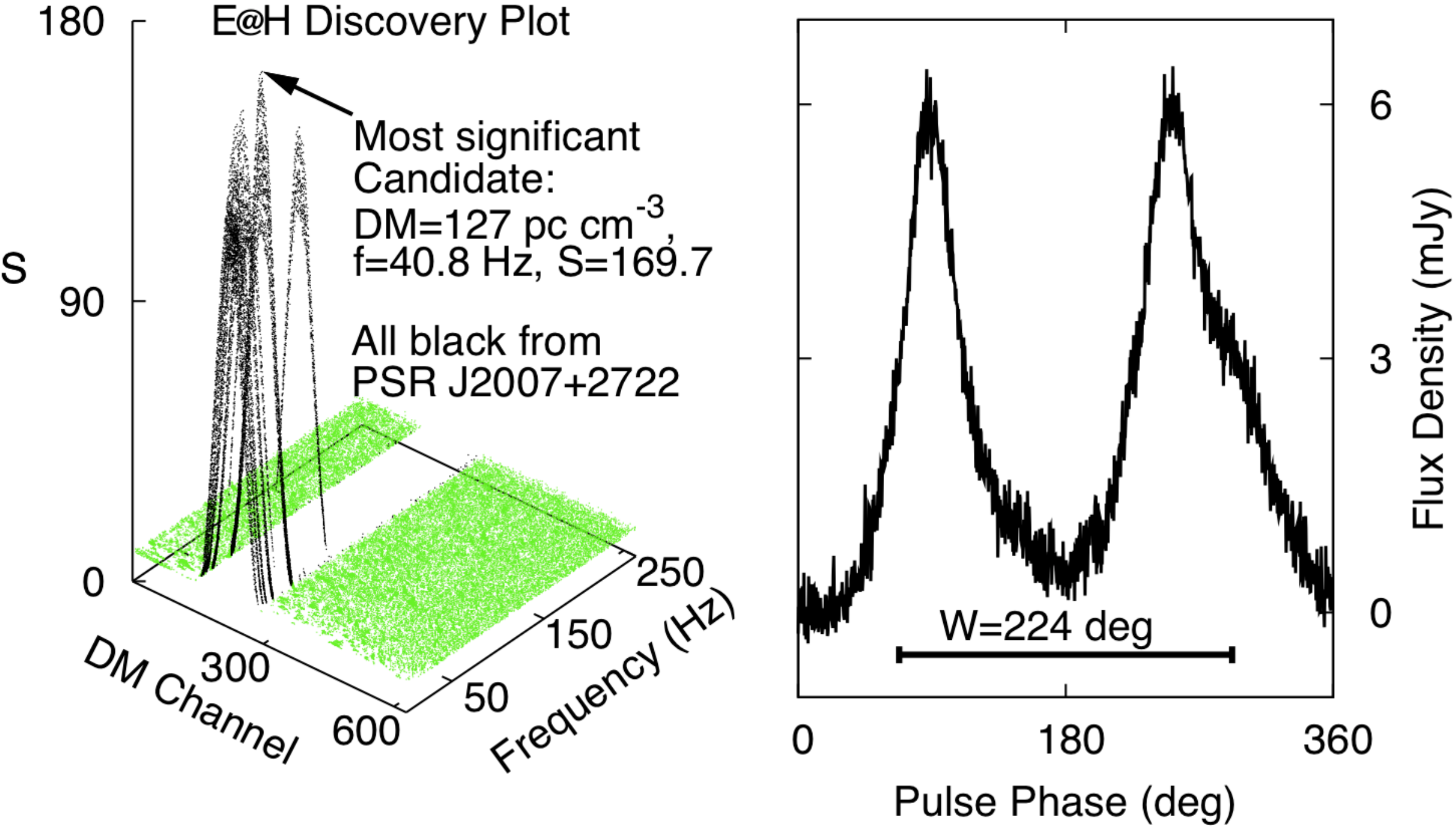}
\caption{{\bf (Left)} significance $\mathcal S$ as a function of DM and spin frequency (all E@H
results for the discovery beam).  {\bf (Right)} the pulse profile at 1.5 GHz (GBT). The bar
illustrates the extent of the pulse.}
\end{figure}

On 11 July, the 24-ms PSR~J2007+2722 was discovered with a significance of $\mathcal
S=169.7$ (Fig.~1).  The pulsar was later re-detected in another PALFA survey
observation.  Follow-up observations were done by the Green Bank
Telescope (GBT, USA), the Lovell Telescope at Jodrell Bank Observatory
(UK), the radio telescope at Effelsberg (Germany), the Westerbork
Synthesis Radio Telescope (WRST, Netherlands), and Arecibo.  
The period-averaged flux density is $2.1$~mJy at 1.5~GHz.  Gridding
observations using Arecibo and WSRT unambiguously associate the pulsar
with a source in an archival VLA C-array observation, having 1.2 mJy flux density at 4.86
GHz, at RA~20$^{\rm h}$07$^{\rm m}$15$^{\rm s}$.77
Dec~27$^\circ$22$'$47$''$.7 (J2000) with uncertainty $\lesssim 1''$.  
The pulsar is not in a supernova remnant or globular cluster and has
no counterpart in X-ray or gamma-ray point-source catalogs.  The DM of
127~pc~cm$^{-3}$ implies a distance of 5.3~kpc\cite{NE2001}.  The full
pulse width between the outer half-maxima is $W \approx
224^\circ$. The wide pulse and initial polarization observations
indicate that the pulsar likely has nearly aligned magnetic and spin
axes.

The pulsar's barycentric spin frequency \cite{tempo} is \freq at MJD 55399.0.  
With the VLA position, the 2010
data give limits $|\dot f| < 3\times 10^{-14}$s$^{-2}$,
magnetic field $B<2.1\times 10^{10}$~G, and spin-down age
$>$21~Myr.  These limits and lack of a
companion mean that J2007+2722 is likely the fastest-spinning
disrupted recycled pulsar yet found \cite{Belczynskietal}.  However we
cannot rule out it having been born with low $B$ \cite{GotthelfHalpern}.  
Either way, PSR~J2007+2722 is a rare, isolated low-$B$ pulsar which
contributes to our understanding of pulsar evolution.

This result demonstrates the capability of ``consumer'' computational power
for realizing discoveries in astronomy and other data-driven science.

We thank Einstein@Home volunteers, who made this discovery possible.
The computers of Chris and Helen Colvin (Ames, Iowa, USA) and Daniel
Gebhardt (Universit\"at Mainz, Musikinformatik, Germany) identified
J2007+2722 with the highest significance.  This work was supported by CFI,
CIFAR, FQRNT, MPG, NAIC, NRAO, NSERC, NSF and STFC; see the
Supporting Online Material for details.

\bibliographystyle{Science}

% break to new page for SOM
\newpage

\section*{Supporting Online Material for ``Pulsar Discovery by Global Volunteer Computing''}

\subsection*{Materials and Methods}
The observations use a seven feed-horn, dual-polarization radio camera at
1.4 GHz (ALFA)\cite{PALFA}. Autocorrelation  spectrometers sum polarizations
and generate spectra over 100 MHz of bandwidth with 256 channels every
64~$\mu s$. 
Data are independently searched for isolated pulsars in
three independent pipelines: (1) the Einstein@home pipeline described in this paper,
searching for isolated or binary pulsars with orbits longer than 11 minutes, (2) a pipeline
at Cornell University's Center for Advanced Computing, and (3) a pipeline using the PRESTO
\cite{PRESTO} package operating at several Pulsar ALFA (PALFA) Consortium member sites, 
searching for isolated or binary pulsars with orbits longer than $\sim$1h.

Timing data were collected from the GBT, Arecibo, Jodrell Bank, and
Effelsberg over a total of 18 days.  Pulse times of arrival were calculated
using standard procedures and analyzed using the tempo software package \cite{TEMPO}.
The timing analysis used the JPL DE405
solar system ephemeris, and times of arrival were referred to local
observatory clock standards.

\subsection*{Acknowledgements}
The Arecibo Observatory is part of the National Astronomy and
Ionosphere Center, which is operated by Cornell University under a
cooperative agreement with the National Science Foundation (NSF).
The National Radio Astronomy Observatory is a facility of the NSF
operated under cooperative agreement by Associated Universities, Inc.
PALFA is supported at Cornell University and Columbia University by
the NSF.
% AEI
The AEI Hannover receives support through a cooperative agreement
between Leibniz Universit\"at Hannover and the Max Planck
Gesellschaft.
% B. Knispel
B.K. thanks the IMPRS on Gravitational Wave Astronomy for its support.
% Allen, Anderson, Papa, Siemens,
We acknowledge NSF support for Einstein@Home through grant
PHY-0555655.
%David Anderson
BOINC is supported by NSF grant OCI-0721124.
% Slavko Bogdanov
S.B. is supported by a Canadian Institute for Advanced
Research (CIFAR) Junior Fellowship.
% Fernando Camilo
F. Camilo acknowledges support from NSF grant AST-0806942.
% Froney Crawford
F. Crawford was supported by grants from Research Corporation and the
Mount Cuba Astronomical Foundation.
%Paul Demorest
P.B.D. is a Jansky Fellow of the National Radio Astronomy Observatory.
% Jason Hessels
J.W.T.H. is a Veni Fellow of The Netherlands Organisation for Scientific -Research (NWO).
%V. Kaspi
V.M.K. is supported by an NSERC Discovery Grant, the Canadian Institute for Advanced
Research, les Fonds qu\'eb\'ecois de la recherche sur la nature et les technologies,
and holds the Lorne Trottier Chair and a Canada Research Chair.
%P. Lazarus
P.L. is supported by a NSERC PGS-M award.
% A. Lyne and B. Stappers
Pulsar research at JBCA and use of the Lovell Telescope is
supported by a Rolling Grant from the UK Science and Technology
Facilities Council.
%Maura and Dunc
M.A.M. and D.R.L. are supported by a WVEPSCOR grant and Cottrell Scholar
Awards. M.A.M. is an Alfred P. Sloan Research Fellow.
% David Nice
We acknowledge support from NSF grant AST 0647820 to Bryn Mawr
College.
% Ingrid Stairs and Marjorie E. Gonzalez
I.H.S. and M.E.G. were supported by an NSERC Discovery Grant and Discovery Accelerator Supplement and by the Canada Foundation for Innovation.
% help from G. Heald
We gratefully acknowledge the help of George Heald (ASTRON) with processing of imaging
data from the Westerbork Synthesis Radio Telescope.
The Westerbork Synthesis Radio Telescope is operated by the Netherlands Institute for Radio Astronomy (ASTRON).

\end{document}